\documentclass[a4paper,11pt]{article}
\pdfoutput=1 

\usepackage{jinstpub} 
\usepackage{graphicx}
\usepackage{subfigure}
\usepackage{lineno}

\title{\boldmath Depleted fully monolithic CMOS pixel detectors using a column based readout architecture for the ATLAS Inner Tracker upgrade}

\author[a,1]{T.~Wang,\note{Corresponding author.}}
\author[b]{M.~Barbero,}
\author[c]{I.~Berdalovic,}
\author[a]{C.~Bespin,}
\author[b]{S.~Bhat,}
\author[b]{P.~Breugnon,}
\author[a]{I.~Caicedo,}
\author[c]{R.~Cardella,}
\author[b]{Z.~Chen,}
\author[d]{Y.~Degerli,}
\author[c]{N. Egidos,}
\author[b]{S.~Godiot,}
\author[d]{F.~Guilloux,}
\author[a]{T.~Hemperek,}
\author[a]{T.~Hirono,}
\author[a]{H.~Kr\"uger,}
\author[c]{T.~Kugathasan,}
\author[a]{F.~H\"ugging,}
\author[c]{C.A.~Marin Tobon,}
\author[a]{K.~Moustakas,}
\author[b]{P.~Pangaud,}
\author[d]{P.~Schwemling,}
\author[c]{H. Pernegger,}
\author[a]{D-L.~Pohl,}
\author[b]{A.~Rozanov,}
\author[a]{P.~Rymaszewski,}
\author[c]{W.~Snoeys,}
\author[a]{and N.~Wermes}

\affiliation[a]{University of Bonn,\\Nussallee 12, Bonn, Germany}
\affiliation[b]{Centre de physique des particules de Marseille,\\163 Avenue de Luminy, Marseille, France}
\affiliation[c]{CERN\\CH-121 Geneve 23, Switzerland}
\affiliation[d]{IRFU, CEA-Saclay\\Gif-sur-Yvette Cedex, 91191 France}

\emailAdd{t.wang@physik.uni-bonn.de}

\abstract{Depleted monolithic active pixel sensors (DMAPS), which exploit high voltage and/or high resistivity add-ons of modern CMOS technologies to achieve substantial depletion in the sensing volume, have proven to have high radiation tolerance towards the requirements of ATLAS in the high-luminosity LHC era.  Depleted fully monolithic CMOS pixels with fast readout architectures are currently being developed as promising candidates for the outer pixel layers of the future ATLAS Inner Tracker, which will be installed during the phase II upgrade of ATLAS around year 2025. In this work, two DMAPS prototype designs, named LF-MonoPix and TJ-MonoPix, are presented. LF-MonoPix was designed and fabricated in the LFoundry 150~nm CMOS technology, and TJ-MonoPix has been designed in the TowerJazz 180~nm CMOS technology. Both chips employ the same readout architecture, i.e. the column drain architecture, whereas different sensor implementation concepts are pursued. The design of the two prototypes will be described. First measurement results for LF-MonoPix will also be shown.
}

\keywords{Depleted monolithic CMOS pixels, particle tracking detectors (solid-state detectors), Front-end electronics for detector readout, VLSI circuit}

\begin{document}
\maketitle
\flushbottom
\section{Introduction}
\label{sec:Intro}

Monolithic active pixel sensors (MAPS) have already matured enough to be used in high energy physics experiments, as high precision tracking and vertexing devices~\cite{HFT_Greiner_2015,ALPIDE_Mager_2016,MuPix_Augustin_2017}. They are now making their path into the high-rate and high-radiation applications, where a depleted sensing volume is mandatory for charge to be collected sufficiently fast by drift~\cite{DMAPS_NW_2015}. CMOS pixels with depleted sensing volume, also labelled as depleted monolithic active pixel sensors (DMAPS), can be achieved by exploiting the high-resistivity and/or high-voltage add-ons of modern CMOS technologies. Many such devices have been reported, showing high radiation tolerance towards the requirements of experiments in the High-Luminosity LHC (HL-LHC) era~\cite{AMSH18_García_2016,CCPD_Irradiation_SubPix_Hirono_2016,LF_Passive_Mandić_2017,Investigator_Pernegger_2017}. Moreover, multiple nested wells, available in many commercial CMOS processes, allow to implement full CMOS electronics inside the pixel (see Section~\ref{sec:SensorConcepts}), and therefore fast readout architectures like hybrid pixels are possible, beyond the rolling shutter readout traditionally used in MAPS. 

As a part of the CMOS program for the future ATLAS Inner Tracker (ITk), DMAPS integrating fast readout architectures are under development. They may serve as a high performance and cost effective option for the outer pixel layers of ITk, replacing the complex and expensive hybrid pixel modules. This work presents two DMAPS prototypes, i.e. LF-MonoPix and TJ-MonoPix. Both of them use a column based readout architecture named column drain readout, similar to the one used in the current ATLAS pixel detector~\cite{FEI3_Ivan_2006}. The LF-MonoPix was designed and fabricated in the LFoundry 150~nm CMOS technology, featuring a large charge collection electrode, usually preferred for high radiation tolerance. In contrast, a small sensing node is used in TJ-MonoPix, where a fully depleted sensitive layer can be expected by profiting from a modified process in the TowerJazz 180~nm CMOS imaging technology~\cite{ModifiedProcess_Snoeys_2017}. The two different sensor concepts will be described in Section~\ref{sec:SensorConcepts}, followed by an overview of the chip design in Section~\ref{sec:Chip}. In Section~\ref{sec:Measurements}, first measurement results for LF-MonoPix will be shown. Section~\ref{sec:Summary} will summarize the work.


\section{Sensor concepts}
\label{sec:SensorConcepts}

The implementation of a depleted CMOS pixel cell can be generally categorized by the fill factor, i.e. the ratio of the sensing electrode area to the total pixel area~\cite{SiliconPixel_NW}. Both large and small fill factor approaches are pursued in this work and implemented as described in the following.

\begin{figure}[!ht]
  \begin{center}
    \subfigure[]{
      \includegraphics[width=0.4\textwidth]{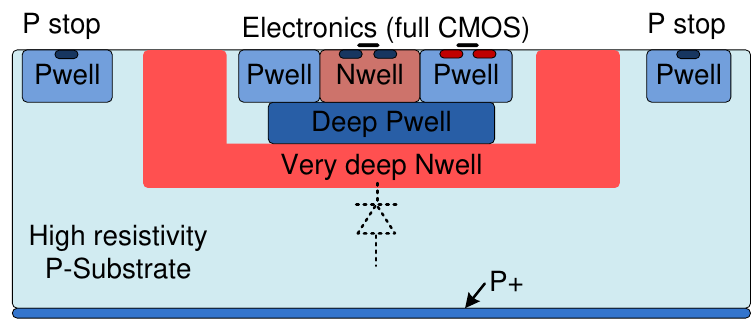}
      \label{fig:LF_Tech}
    }
    \qquad
    \subfigure[]{
      \includegraphics[width=0.37\textwidth]{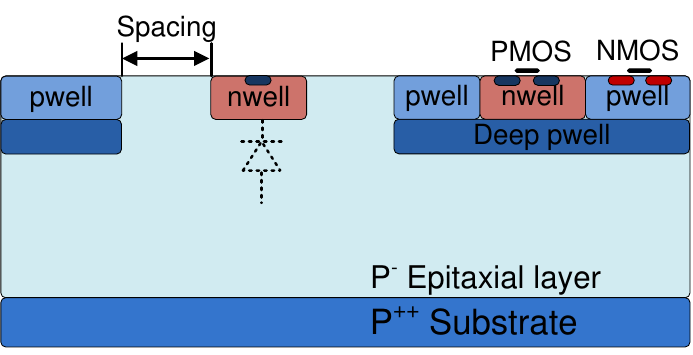}
      \label{fig:TJ_Tech}
    }
    \caption[]{Cross section view of one pixel cell implemented \subref{fig:LF_Tech} in the LFoundry 150 nm CMOS technology featuring a large charge collection electrode and \subref{fig:TJ_Tech} in the TowerJazz 180 nm CMOS technology featuring a small sensing node.} 
    \label{fig:Tech}
  \end{center}
\end{figure}

\paragraph{LF-MonoPix}
Figure~\ref{fig:LF_Tech} shows the cross section view of a pixel cell in LF-MonoPix. The sensing volume is a high resistivity P-substrate (> 2~k$\Omega\cdot$cm). The charge collection node is formed by a very deep N-well, which encloses the in-pixel electronics. Full CMOS circuitry is possible because of the isolation between the N-well, hosting the PMOS transistors, and the charge collection N-well, by a deep P-well. High charge collection efficiency after radiation damage can be expected from a large charge collection electrode, together with the possibility of applying back-side reverse bias to achieve more uniform drift field. Measurement of a previous prototype shows that a depletion depth over 100~$\mu$m is achievable with such sensor design after particle fluence of 10$^{15} $~n$_{eq}$/cm$^{2}$~\cite{LF_Mandić_2017}, the expected NIEL radiation fluence for the ITk outer pixel layers. However, it may suffer from large detector capacitance, e.g. $\sim $~400~fF, especially when complex in-pixel logic is needed to implement a fast readout architecture~\cite{SiliconPixel_NW,MonoPix_TWang_2017}. The large detector capacitance will slow down the pre-amplifier and increase the noise, which can only be compensated by increasing the power. Moreover, such sensors are sensitive to the cross talk introduced by the in-pixel electronics to the sensing node, and many design efforts are required to mitigate this issue~\cite{MonoPix_TWang_2017}.

\paragraph{TJ-MonoPix}
 
As depicted in Figure~\ref{fig:TJ_Tech}, TJ-MonoPix uses a high-resistivity (>~1~k$\Omega\cdot$cm) P-type epitaxial layer as the sensitive layer, located on top of a highly P-doped wafer substrate. The typical thickness of the epitaxial layer is between 18~$\mu$m and 40~$\mu$m. The charge collection electrode is created by a small N-well (e.g. several $\mu$m$^{2}$). The deep P-well is used to shield the N-wells that host the PMOS transistors, which would otherwise work as competing charge collection nodes. The main motivation of using a small sensing diode is its very small detector capacitance (e.g. < 5 fF), which allows to minimize the analog power consumption with given analog performance~\cite{MAPS_Snoeys_2014}. The use of small collection node in high radiation environment is also encouraged by recent R\&D progress to improve the sensor depletion by using a modified process~\cite{ModifiedProcess_Snoeys_2017}. Despite the small charge collection electrode, a test chip called Investigator, fabricated with the modified process, showed uniform efficiency for 25~$\mu$m and 30~$\mu$m square pixels, even after fluence of 10$^{15} $~n$_{eq}$/cm$^{2}$~\cite{Investigator_Pernegger_2017}.


\section{Chip overview}
\label{sec:Chip}

LF-MonoPix is the first fully monolithic prototype of a DMAPS series implemented in the LFoundry technology~\cite{CCPD_FirstResults_2016,LFCPIX_Yavuz_2017}. Its design has significant inputs from its ancestor LF-CPIX~\cite{LFCPIX_Yavuz_2017}, e.g. pre-amplifier, guard ring structure, chip floor plan. DMAPS development in the TowerJazz technology follows the successful experience of the ALPIDE chip for the ALICE-ITS upgrade~\cite{ALPIDE_Mager_2016}. The TJ-MonoPix is part of the TowerJazz DMAPS R\&D, aimed to translate the well understood column drain architecture, based on the experience of LF-MonoPix, into the TowerJazz sensor design with small charge collection node. The two chips are large scale demonstrator chips. Some main parameters for these designs are listed in Table~\ref{tab:Chip}. 

\begin{table}[htbp]
\centering
\caption{\label{tab:Chip} Main parameters of LF-MonoPix and TJ-MonoPix.}
\smallskip
\begin{tabular}{|l|c|c|}
\hline
 Parameter & LF-MonoPix & TJ-MonoPix\\
\hline
CMOS tech. & LFoundry 150~nm & TowerJazz 180~nm\\
\hline
Sensor concept & Large fill factor & Small fill factor\\
\hline
Chip size & $\sim$~1~$\times$~1~cm$^{2}$ & $\sim$~1~$\times$~2~cm$^{2}$\\
\hline
Pixel pitch & 50~$\times$~250~$\mu$m$^{2}$ & 36~$\times$~40~$\mu$m$^{2}$\\
\hline
Pixel matrix & 129~$\times$~36 & 224~$\times$~448\\
\hline
Static current/pixel & $\sim$~20~$\mu$A & <~1~$\mu$A\\
\hline
Output data link & 1~$\times$~LVDS@160~MHz & 4~$\times$~CMOS@40~MHz\\
\hline
\end{tabular}
\end{table}

\subsection{Architecture}
\label{sec:Arch}

\begin{figure}[htbp]
	\centering
	\includegraphics[width=.9\textwidth]{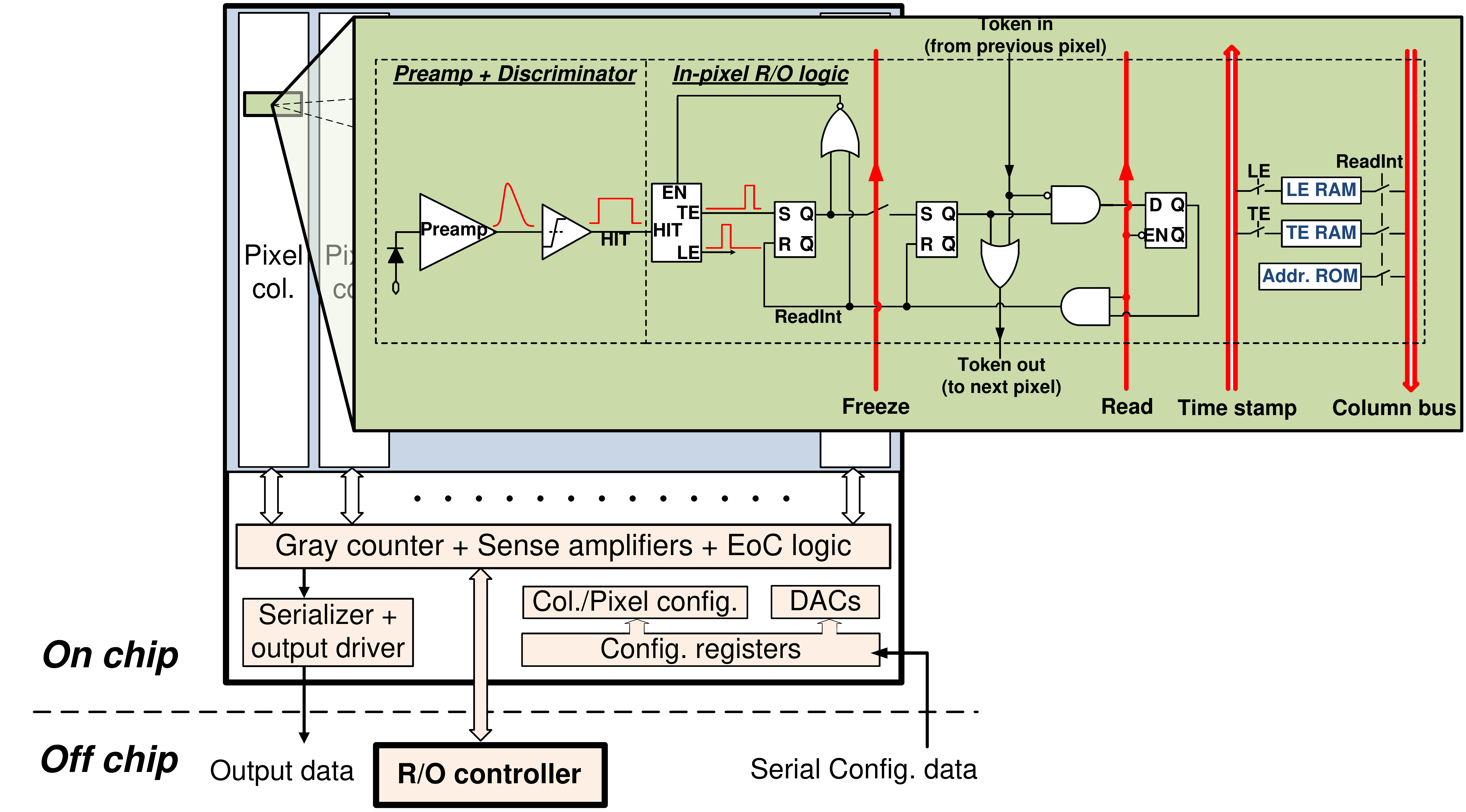}
	\caption{The chip architecture of the two prototypes in this work.} 
	\label{fig:chip_arch}
\end{figure}

Though having different layout floor plans, both LF-MonoPix and TJ-MonoPix follow the same architecture as depicted in Figure~\ref{fig:chip_arch}. 

The readout chain starts with the pixel front-end (FE) circuit, i.e. the pre-amplifier and the discriminator. The discriminator fires when the analog signal pulse crosses its threshold, and the output holds until the analog pulse falls below the threshold. Coarse analog information can be obtained by measuring the width of the digital pulse at the discriminator output, a technique called time over threshold (ToT). Two gray encoded time stamps, corresponding to the leading edge (LE) and trailing edge (TE) of the discriminator output, are written into the two in-pixel RAM cells to record the hit time and pulse width.  The pixel readout is arbitrated by a token propagation, with the topmost pixel having the highest readout priority. The pixel column interfaces with the End-of-Column (EoC) circuitry, which includes the gray counter running at 40~MHz to generate the time stamp, the sense amplifiers to receive the column data, and the EoC logic to perform the column-level readout priority scan and transmit data to the serial link. In these first prototypes, for design simplicity, the readout controller, which controls the readout sequence, is implemented in an FPGA. Triggering buffers are not included either, hence all the hit data is transmitted through the serial link off chip.

It is noted that a second readout concept was implemented in LF-MonoPix, where the pixel contains only the FE circuit. The digital block is placed at the column end, with one to one connection to the corresponding pixel. This was mainly motivated by reducing the potential cross talk to the sensing node introduced by the digital logic, but more complex wiring in the matrix layout was needed.

\subsection{Pixel}
\label{sec:Pixel}

The different sensor concepts of the two prototype chips have led to different pixel designs in terms of sensor geometry and FE circuit.

\begin{figure}[htbp]
  \centering
  \includegraphics[width=.58\textwidth]{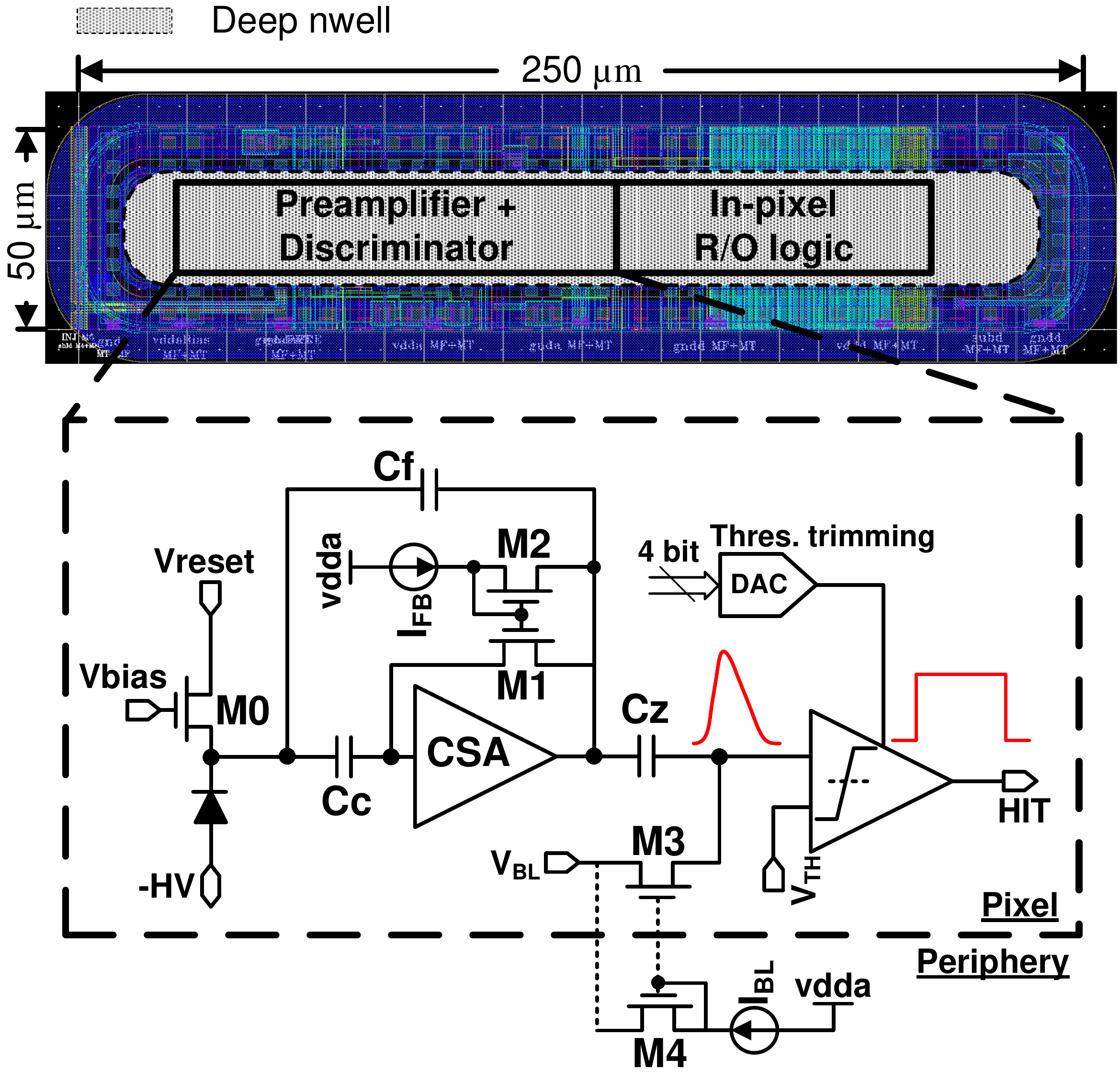}
  \caption{Layout of a typical pixel cell in LF-MonoPix (upper) and schematic of its FE circuit (lower).} 
  \label{fig:LF_Pix}
\end{figure}

\paragraph{LF-MonoPix}
The layout of a typical pixel in LF-MonoPix is shown in Figure~\ref{fig:LF_Pix} (upper). The pixel size is 50~$\times$~250~$\mu$m$^{2}$. The charge collection well is represented by the shaded area, giving a fill factor about 55\%. The schematic of the FE circuit is also shown in Figure~\ref{fig:LF_Pix} (lower). The pre-amplification stage is a typical charge sensitive amplifier (CSA), AC coupled to the sensor. The signal charge is integrated on the feedback capacitor $C_{f}$. The DC feedback, mainly composed of the current mirror M1 and M2, stabilizes the operation point and continuously discharges the integrated signal. Such a DC feedback allows for high feedback resistance, regardless the discharge current which is defined by $I_{FB}$~\cite{FE_Blanquart_1997}. The output of the CSA is sent to a discriminator, whose threshold can be trimmed by a 4-bit in-pixel DAC. Thanks to the AC coupling between the CSA and discriminator, the baseline of the discriminator input is set to $V_{BL}$ via the MOS resistor M3, independent on the DC level of the CSA. The biasing of M3 is set globally by an on-chip current DAC adjusting the current $I_{BL}$. 

\begin{figure}[htbp]
  \centering
  \includegraphics[width=.9\textwidth]{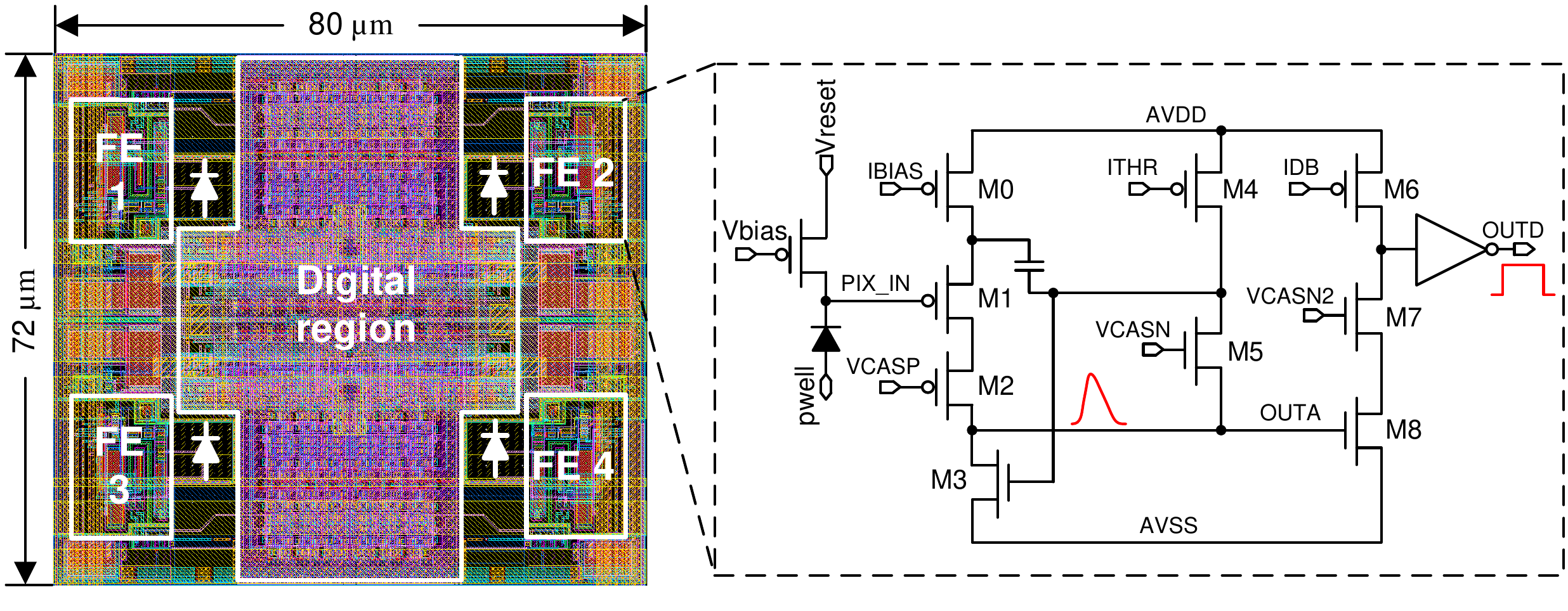}
  \caption{Layout of four neighboring pixels (left) and schematic of the FE circuit (right) of TJ-MonoPix.} 
  \label{fig:TJ_Pix}
\end{figure}

\paragraph{TJ-MonoPix}
The layout of 4 neighboring pixels in TJ-MonoPix is shown in  Figure~\ref{fig:TJ_Pix} (left). Each pixel has an area of 36~$\times$~40~$\mu$m$^{2}$. The charge collection well is located in the center of the pixel, and it has an octagon shape with a diameter of 2~$\mu$m. The spacing between the charge collection node and its nearest P-well is 3~$\mu$m (refer to Figure~\ref{fig:TJ_Tech}). The estimated detector capacitance is less than 5~fF, almost two orders of magnitude smaller than LF-MonoPix. The small detector capacitance allows the use of a very compact and low power FE circuit derived from the ALPIDE chip~\cite{AlpideFE_Kim_2016}, the schematic of which is sketched in Figure~\ref{fig:TJ_Pix} (right). The signal charge is integrated on the capacitance of the input node $PIX\_IN$. It is noted that the low capacitance leads to large voltage excursion at the input node. The resulting voltage signal is then amplified and shaped by the FE circuit, generating a voltage pulse at the node $OUTA$. The charge-to-voltage conversion gain seen at $OUTA$ can be as high as several mV/e$^{-}$. Such a high gain makes it possible to use a simple inverting stage (M6~-~M8 in Figure~\ref{fig:TJ_Pix}) as the discriminator without threshold trimming. 


\section{First results for LF-MonoPix}
\label{sec:Measurements}

The LF-MonoPix wafers were back from the foundry in March 2017, and the chip is fully functional. Extensive characterization is currently in progress, both in lab and in beams. This section shows some first laboratory measurement results. 

\begin{figure}[htbp]
  \centering
  \includegraphics[width=.42\textwidth]{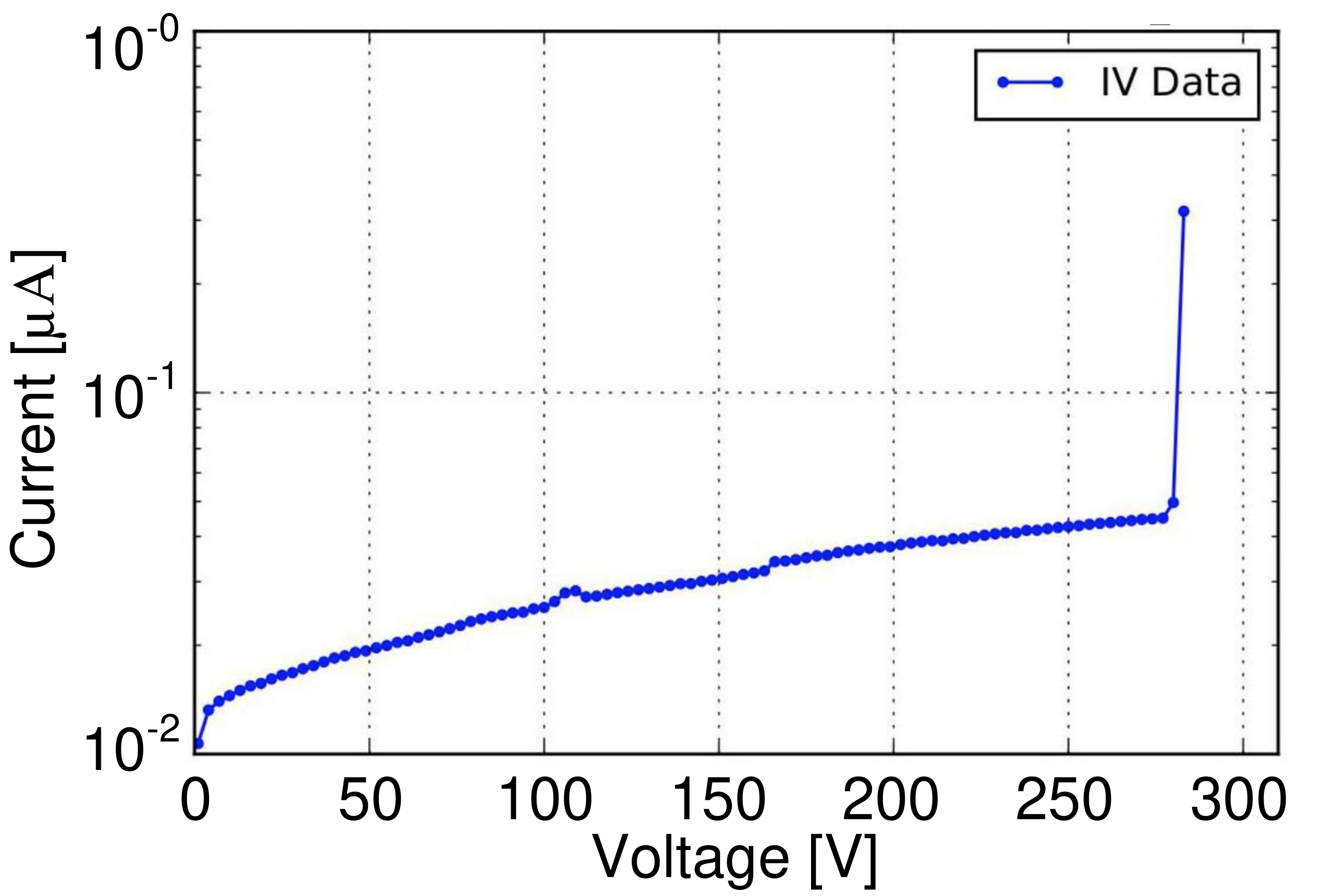}
  \caption{I-V curve of LF-MonoPix measured at room temperature. The voltage value refers to the reverse bias voltage.} 
  \label{fig:IV_Curve}
\end{figure}

\paragraph{I-V curve}
The sensor depletion can be enhanced by a high bias voltage. The highest voltage that can be applied is limited by the sensor breakdown behavior. Therefore, the sensor leakage current was measured as a function of reverse bias voltage. Figure~\ref{fig:IV_Curve} shows the I-V curve of LF-MonoPix measured at room temperature. 
The breakdown was measured at -280~V, an improvement over previous chip generation in the same technology~\cite{CCPD_FirstResults_2016}, thanks to a new guard ring structure~\cite{LFCPIX_Yavuz_2017}. Such high breakdown voltage is likely to ensure a sufficient depletion after irradiation~\cite{LF_Passive_Mandić_2017,LF_Mandić_2017}.

\begin{figure}[!ht]
  \begin{center}
    \subfigure[]{
      \includegraphics[width=0.48\textwidth]{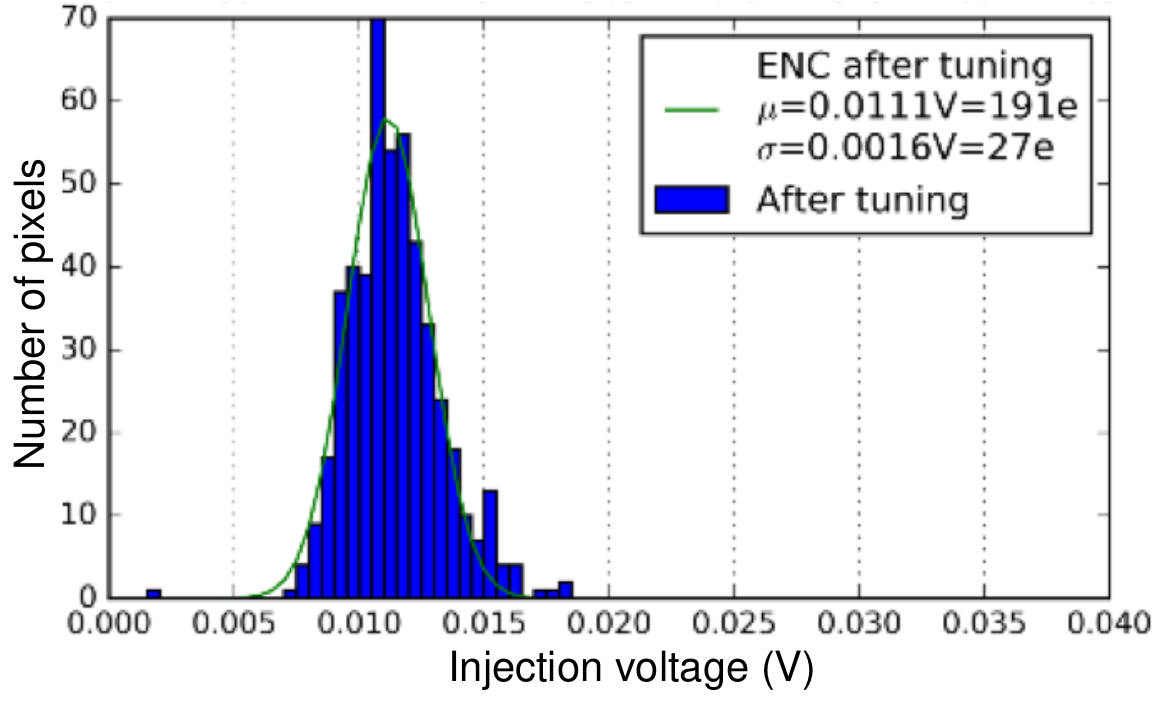}
      \label{fig:Noise_24_27}
    }
    \subfigure[]{
      \includegraphics[width=0.48\textwidth]{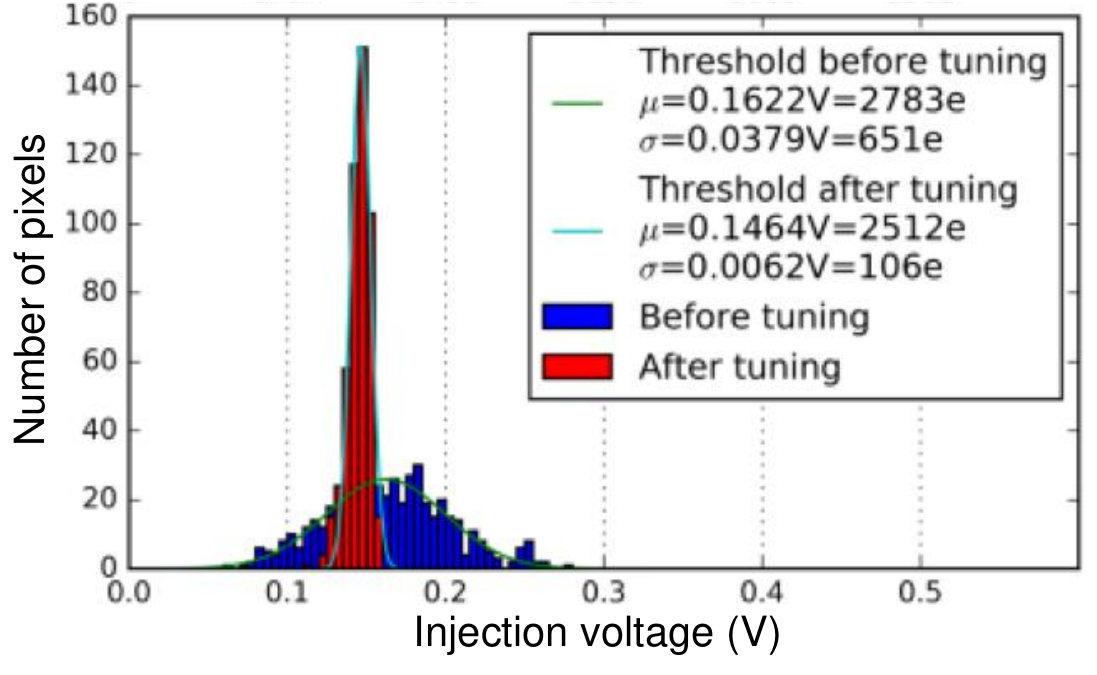}
      \label{fig:Th_24_27}
    }
    \caption[]{\subref{fig:Noise_24_27} Noise distribution and \subref{fig:Th_24_27} Threshold distribution before (blue) and after (red) tuning for one pixel variant with full in-pixel readout logic and using a CMOS input pre-amplifier.} 
    \label{fig:Noise_Th}
  \end{center}
\end{figure}

\paragraph{Noise and threshold distribution}
The noise performance and threshold distribution were studied by scanning an external injection voltage and recording the sensor response. The injected signal was calibrated by an $^{241}$Am radioactive source and X-ray fluoresce of Terbium. Figure~\ref{fig:Noise_Th} shows the noise and threshold distribution of four pixel columns, composed of one pixel variant with full in-pixel readout logic. Its pre-amplifier has complementary input transistors as described in~\cite{LFCPIX_Yavuz_2017}. The average noise value is 191~e$^{-}$, with dispersion of 27~e$^{-}$. The threshold was tuned to around 2500~e$^{-}$, and the dispersion is about 100~e$^{-}$. It is noted that the threshold shown here should not be regarded as the minimum operation threshold. A lower value may be achieved by improving the tuning algorithm. Systematic study on the performance of different pixel variants is still ongoing, which will allow us to choose the best design in terms of pre-amplifier, discriminator, layout scheme and readout concept. 

\section{Summary}
\label{sec:Summary}

Two large scale DMAPS prototypes are presented in this work, both having the column drain readout electronics integrated on the sensor substrate. The goal is to demonstrate the feasibility of using fully monolithic CMOS pixels in the outer pixel layers of the future ATLAS ITk.

The LF-MonoPix, fabricated in the LFoundry 150~nm CMOS technology, employs the large fill factor sensor concept, as a safe choice for radiation hardness. Preliminary test results on the LF-MonoPix IC (Integrated Circuit) show that the chip is fully functional with a high break down voltage of -280~V. First measurement of one pixel variant having full in-pixel digital logic showed a tunable threshold of $\sim$~2500~e$^{-}$ and a noise value of $\sim$~200~e$^{-}$, which are comparable to the current ATLAS pixel detector~\cite{ATLASPixelPerformance_Heim_2014}. Two test beam campaigns have been recently performed, one using the electron beam at ELSA (University of Bonn), the other with pions at CERN SPS. The efficiency and timing analysis are ongoing. Chip samples irradiated with neutrons and protons, with dose levels up to 2~$\times$~10$^{15}$~n$_{eq}$/cm$^{2}$ and 150~MRad respectively, will also be characterized to study the irradiation performance. 

The TJ-MonoPix, which has been recently submitted, is implemented in the TowerJazz 180~nm CMOS technology. By using a novel modified process, full depletion in the sensitive layer can be expected even with a small sensing electrode. Its small detector capacitance brings the benefits of low power (<~2~$\mu$W/pixel) and small pixel (< 50~$\mu$m~$\times$~50~$\mu$m). The TJ-MonoPix demonstrator is expected to be back from the foundry at the end of 2017. 

\acknowledgments

This work is supported by the H2020 project AIDA-2020, GA no. 654168, and by the H2020 project STREAM, GA no. 675587

\end{document}